\documentclass[aps,prc,twocolumn,superscriptaddress,showpacs,floatfix]{revtex4}
\usepackage{amssymb}
\usepackage{enumitem}
\usepackage{graphicx}%
\usepackage{amsmath}%
\usepackage{amsfonts} 
\usepackage{subfigure}

\def\scr{\rm\scriptscriptstyle }

\begin{document}

\title{Approximate transmission coefficients in heavy ion fusion}

\author{A. J. Toubiana} 
	\email{ajtoubiana@gmail.com}
	\affiliation{Departamento de Engenharia Nuclear, Escola Polit\'ecnica,  Universidade Federal do Rio de Janeiro,
	C.P. 68529, 21941-909, Rio de Janeiro, RJ, Brazil
	 }
	\affiliation{\'Ecole CentraleSup\'elec, Grande Voie des Vignes, Ch\^atenay-Malabry, 92 295, France}
\author{L.F. Canto}
	\email{canto@if.ufrj.br}
	\affiliation{Instituto de F\'{\i}sica, Universidade Federal do Rio de Janeiro, CP 68528, 21941-972, Rio de Janeiro, RJ, Brazil}
	\affiliation{Instituto de F\'{\i}sica, Universidade Federal Fluminense, Av. Litoranea s/n, Gragoat\'{a}, Niter\'{o}i, R.J., 24210-340, Brazil}
\author{M. S. Hussein}
	\email{hussein@if.usp.br}
	\affiliation{Departamento de F\'isica Matem\'atica, Instituto de F\'isica, Universidade de S\~ao Paulo, C.P. 66318, 05314-970, S\~ao Paulo, SP, Brazil}
	\affiliation{Instituto de Estudos Avan\c{c}ados, Universidade de S\~ao Paulo, C.P. 72012, 05508-970, 
		S\~ao Paulo, SP, Brazil}
	\affiliation{Departamento de F\'{i}sica, Instituto Tecnol\'{o}gico de Aeron\'{a}utica, CTA, S\~{a}o Jos\'{e} dos Campos, S\~ao Paulo, SP, Brazil}

\begin{abstract}
In this paper we revisit the one-dimensional tunnelling problem. We consider different approximations for the transmission through the Coulomb
barrier in heavy ion collisions at near-barrier energies. First, we discuss approximations of the barrier shape by functional forms where the
transmission coefficient is known analytically. Then, we consider Kemble's approximation for the transmission coefficient. We show how this
approximation can be extended to above-barrier energies by performing the analytical continuation of the radial coordinate to the complex plane. 
We investigate the validity of the different approximations considered in this paper by comparing their predictions for transmission coefficients
and cross sections of three heavy  ion systems with the corresponding quantum mechanical results. 
\end{abstract}

\maketitle

%%%%%%%%%%%%%%%%%%%%%%%%%%%%%%%%%%%%%%%%%%%%%%%%%%%%%%%%%%%%%%%%%%%%%%%%%%%%%

\section{Introduction}

%%%%%%%%%%%%%%%%%%%%%%%%%%%%%%%%%%%%%%%%%%%%%%%%%%%%%%%%%%%%%%%%%%%%%%%%%%%%%

The transmission through a potential barrier is a remarkable feature of quantum mechanics. Since Gamow's theory of alpha decay,
in 1928, it has been used to explain a variety of new phenomena. In particular, the transmission coefficients through the Coulomb 
barrier has been an important ingredient in calculations of heavy ion fusion cross sections along the last few decades. In 
single-channel descriptions of heavy ion scattering, fusion is frequently simulated by a strong imaginary potential acting in the inner side of the 
Coulomb barrier. Since the fraction of the incident current that reaches the strong absorption region is fully absorbed, the fusion probability 
at each partial-wave can be approximated by the transmission coefficient through the Coulomb+centrifugal barriers. \\

Although the transmission coefficients through an arbitrary potential barrier can be evaluated by numerical procedures, 
it is convenient to have analytical expressions. This is possible in some particular cases, like the parabolic 
barrier~\cite{FHW59} and the Morse~\cite{Ahm91} barrier. Parabolic barriers have been used for decades in fusion reactions
of heavy ions~\cite{HiW53,Won73,CGD06,CGD15}. In this case, the barrier at each partial wave is approximated by a 
parabola, with properly chosen parameters, and the transmission coefficients are given by an analytical expression involving
these parameters. However, although this approximation is very good near the barrier radius, it becomes progressively worse 
as the radial distance increases. The reason for this shortcoming is that the parabola is symmetric around the barrier
radius, whereas the actual barrier is highly asymmetric. The latter falls off very slowly at large distances (like $1/r$), while 
the former decreases rapidly. This exerts great influence on the 
transmission coefficient at energies well below the barrier, which are very sensitive to the potential at large distances. Thus, 
the parabolic approximation cannot be used to evaluate transmission coefficients in this energy region. The situation is better
if one approximates the Coulomb barrier by a Morse function. Although it leads to a more complicated analytical expression,
this barrier has the advantage of being asymmetric. Similarly to the Coulomb barrier, it falls off slowly at the tail, and rapidly 
in the inner region. \\

Owing to the short wavelengths involved in heavy ion collisions the transmission coefficients are frequently evaluated by 
semiclassical approximations, like the WKB (for a recent review, see Ref.~\cite{HaT12}). However, although this approximation 
has been very successful at energies well below the Coulomb barrier, it fails at energies near the Coulomb barrier and above. 
Kemble~\cite{Kem35} derived an improved version of the WKB approximation that remains valid at energies just below the 
Coulomb barrier. Further, he suggested that his expression for the transmission coefficient could be extended to energies above 
the barrier  through an analytical continuation of the radial variable to the complex plane. In a previous work~\cite{TCH17}, we 
followed this procedure to study Kemble's approximation for a typical heavy ion potential, below and above the barrier. We
concluded that the transmission coefficient and cross sections evaluated in this way were in good agreement with their quantum 
mechanical counterparts. \\

The present work reports a detailed study of parabolic and Morse approximations for the Coulomb\,+\,centrifugal barriers, and 
the resulting transmission coefficients and fusion cross sections. It investigates also the use of Kemble's approximation for
the transmission coefficients through these barriers, at energies above and below the Coulomb barrier. We show that Kemble's 
approximation in these cases becomes exact, independently of the collision energy. The paper is organized as follows.
In Sec.~II, we present a brief description of the single-channel approach to the fusion cross section in heavy ion scattering.
In Sec. III, the main section of this work, we discuss approximate calculations of transmission coefficients and fusion cross
sections. We consider the approximation of the barrier by a parabola and by a Morse function, and different versions of the
WKB approximation for the transmission coefficients. We discuss also the Wong formula for the fusion cross section and
recent improvements to it. In Sec.~IV, we apply the approximations of the previous section to the calculation of
transmission coefficients and fusion cross sections. We consider a few light and medium mass systems, namely 
$^4$He + $^{16}$O, $^{12}$C + $^{16}$O and $^{16}$O + $^{208}$Pb. It is expected that the accuracy of these 
approximations improves with the mass of the system~\cite{CGD06}. Finally, in Sec. V, we present the conclusions of 
the present work.

%%%%%%%%%%%%%%%%%%%%%%%%%%%%%%%%%%%%%%%%%%%%%%%%%%%%%%%%%%%%%%%%%%%%%%%%%%%%%

\section{Fusion reactions in heavy ion collisions}

%%%%%%%%%%%%%%%%%%%%%%%%%%%%%%%%%%%%%%%%%%%%%%%%%%%%%%%%%%%%%%%%%%%%%%%%%%%%%

In  single-channel descriptions of heavy ion scattering the fusion process is usually simulated by an imaginary potential. This potential
is very strong and has a short range, so that it acts exclusively in the inner region of the potential barrier, resulting from nuclear 
attraction plus Coulomb repulsion. The scattering wave function is expanded in partial waves, leading to a radial equation for each 
angular momentum. The real part of the potential, appearing in the radial equations, can be written as
\begin{equation}
V_l(r) = V_{\scr C}(r) + V_{\scr N}(r) + \frac{\hbar^2}{2\mu\, r^2}\ l(l+1),
\label{Veff}
\end{equation}
where the Coulomb interaction between the finite nuclei is usually approximated by the expression
\begin{eqnarray}
V_{\scr C}(r) &=&  \frac{ Z_{\scr P} Z_{\scr T}\,e^2}{r},
\qquad\qquad\qquad\ \ \  \ {\rm for\ } r\ge R_{\scr C}, \label{Vcou}
\\
                     &=& \frac{Z_{\scr P} Z_{\scr T}\,e^2}{2\,R_{\scr C}}\ \left( 3- \frac{r^2}{R_{\scr C}^2} \right), \label{Vcou-1}
                     \qquad {\rm for\ } r < R_{\scr C},
                     \label{Vcou-2}
\end{eqnarray}
with
\[
R_{\scr C} = r_{{\scr 0 C}}\, \left( A_{\scr P}^{\scr 1/3}+ A_{\scr T}^{\scr 1/3}\right).
\]
In the above equations, $Z_{\scr P}$ ($A_{\scr P}$) and $Z_{\scr T}$ ($A_{\scr T}$) are respectively the projectile's and target's atomic
(mass) numbers, and we take $r_{0{\scr C}}\simeq 1$ fm.

\medskip

Different procedures have been proposed to determine the nuclear interaction between two heavy ions (see, e.g. \cite{CaH13} 
and references therein).  Among them, the double folding model~\cite{SaL79} is a systematic procedure that has the advantage 
of being applicable to any heavy ion system. In this model, the potential is given by a multi-dimensional integral involving the
densities of the two nuclei and a realistic nucleon-nucleon interaction. On the other hand, this model has the unpleasant feature
of requiring the evaluation of a rather complicate integral. To avoid this problem, Aky\"uz and Winther~\cite{BrW91} proposed 
a simplified version of the double-folding model. They evaluated the folding integral for a large number of systems, and 
fitted the resulting potential by the Woods-Saxon (WS) function,
 \begin{equation}
 V_{\scr N}(r) = \frac{V_{\scr 0}}  { 1+\exp\left[  \left( r - R_{\scr 0} \right)/a_{\scr 0}  \right]  },
 \label{Vnuc-WS}
 \end{equation}
with $R_{\scr 0} = r_{\scr 0}\,  \left( A_{\scr P}^{\scr 1/3}+ A_{\scr T}^{\scr 1/3}\right)$. The parameters $V_{\scr 0}, r_{\scr 0}$ 
and $a_{\scr 0}$ were then given by analytical expressions of the mass numbers of the collision partners. 
This potential is adopted throughout the present work. The values of the WS parameters for the systems studied here
are given in Table~\ref{tab-WSparam}.\\

\begin{table}
\caption{
Strengths, radius parameters and diffusivities in the Woods-Saxon parametrization of the Aky\"uz-Winther interaction for the 
systems studied in the present paper. 
}
\centering
\begin{tabular} [c] {lccc}
%\begin{tabular} [c] {llll}
\hline \\
%& & \\ 
System:   &   $^4$He + $^{16}$O\ \ \ \ \ \ &   $^{12}$C + $^{16}$O\ \ \ \    & $^{16}$O + $^{208}$Pb   \\ 
 \hline \\
    $V_{\scr 0}$ (MeV) $\qquad$ \ \ & -29.64  & -39.47 & -64.97  \\
    $r_{\scr 0}$ (fm)     \ \ \ \    \ & 1.156  & 1.163 & 1.179  \\
    $a_{\scr 0}$ (fm)      \ \ \ \   \ \ & 0.5535  & 0.5928 & 0.6576  \\

    \hline 
\end{tabular}
\label{tab-WSparam}
\end{table}

Usually, the loss of the incident flux to the fusion channel is simulated by a short range imaginary potential. The radial equation 
is then solved numerically, starting from $r=0$, and the elastic scattering and the fusion cross sections are determined from 
the $l$ components of the S-matrix, $S_l$, given by the asymptotic form of the radial wave function. 

\medskip

An equivalent method to simulate the fusion process is the {\it Ingoing Wave Boundary Condition} (IWBC).
In this method the potential is real but the integration of the radial equation does not start at the origin. It starts at some radial 
distance in the inner region of the barrier (usually the minimum of the total potential), where the radial wave function is assumed to 
have a purely ingoing behaviour. The initial values of the wave function and of its derivative are then evaluated by the WKB 
approximation. We adopt this procedure in the present work.\\

The fusion cross section is then given by the partial-wave sum
 \begin{equation}
 \sigma_{\scr F} = \frac{\pi}{k^2}\ \sum_{l=0}^\infty \left( 2l+1\right) \ P^{\scr F}_l,
 \label{lsumsigF}
 \end{equation}
with the fusion (absorption) probability,
 \begin{equation}
P^{\scr F}_l = 1-\left| S_l \right|^2.
 \label{PFl}
 \end{equation}

 \medskip
 
 Since the wave function inside the barrier is totally absorbed, the fusion probability must be very close to the probability that
 the incident current reaches the point of total absorption. Thus, it can be approximated by the tunnelling probability through
 the corresponding $l$-dependent barrier. That is,
\begin{equation}
P^{\scr F}_l \simeq T_l .
\label{PlTl}
\end{equation}
We use this approximation to derive the fusion cross sections from the transmission coefficients of our approximate calculations.

%%%%%%%%%%%%%%%%%%%%%%%%%%%%%%%%%%%%%%%%%%%%%%%%%%%%%%%%%%%%%%%%%%%%%%

\section{Approximate transmission coefficients and fusion cross sections}

%%%%%%%%%%%%%%%%%%%%%%%%%%%%%%%%%%%%%%%%%%%%%%%%%%%%%%%%%%%%%%%%%%%%%%

Now we discuss different approximations for the transmission coefficients mentioned in the previous section. We consider 
approximations of the Coulomb barrier itself and discuss the use of different versions 
of the WKB approximation in the calculation of transmission coefficients. We discuss also Wong's approximation for the fusion 
cross section, which is widely used in the study of heavy ion fusion.

\subsection{Approximations of the potential barriers}

The transmission coefficient for some particular barriers can be evaluated analytically. This is the case of the 
parabolic~\cite{FHW59} and the Morse~\cite{Ahm91} barriers. These results can be used in calculations of fusion cross 
sections in heavy ion collisions. For this purpose, one approximates the potential barriers of $V_l(r)$ by parabolae or 
by Morse functions, with properly chosen parameters. To simplify the discussion, we consider only S-waves. Other 
angular momenta can be handled similarly.

\subsubsection{Parabolic barrier}

Let us consider a parabolic barrier written as
\begin{equation}
V(r) = V_{\scr B} - \frac{1}{2}\,\mu \omega^2\ \left( r-R_{\scr B} \right)^2.
\label{parab-bar}
\end{equation}
It corresponds to an inverted Harmonic Oscillator with the maximum at $r=R_{\scr B}$, with the value $V\left( r= R_{\scr B}\right) = 
V_{\scr B}$. The barrier curvature parameter, $\hbar\omega$, is related to the second derivative of the total potential at 
$r=R_{\scr B}$ by the equation,
\begin{equation*}
\hbar\omega = \sqrt{
\frac{ - \, \hbar^2\,V^{\prime\prime}\left( R_{\scr B}\right)} {\mu}}.
%\label{hbaromega}
\end{equation*}

\medskip

The transmission coefficient through this barrier, known in the literature as the {\it Hill-Wheeler transmission coefficient}, 
can be written as
\begin{equation}
T_0^{\scr HW} (E)= \frac{1}{1+\exp\left[2\,\Phi^{\scr HW}(E) \right]},
\label{T0HW}
\end{equation}
with 
\begin{equation}
\Phi^{\scr HW}(E)= \frac{\pi}{\hbar\omega}\ \left(V_{\scr B} - E\right).
\label{PhiHW}
\end{equation}
Note that the transmission coefficient of Eqs.~(\ref{T0HW}) and (\ref{PhiHW}) is exact.

\subsubsection{The Morse barrier}
 
The Morse barrier is given by the expression
\begin{equation}
V_{\scr M}(r) = V_{\scr B}\ \Big[ 2\,e^{-\left(r-R_{\scr B}\right)/a_{\scr M}} - 
\,e^{-2\,\left( r-R_{\scr B} \right)/a_{\scr M}}  \Big],
\label{Vmorse}
\end{equation}
where $R_{\scr B}$ and $V_{\scr B}$ are respectively the radius and the height of the barrier. Above, $a_{\scr M}$ is the Morse 
parameter, which is related to the second derivative of the potential at the barrier radius, 
$V^{\prime\prime}\left( R_{\scr B} \right)$, by the expression,
\begin{equation*}
a_{\scr M} = \sqrt{
\frac{2\,V_{\scr B}}{-V^{\prime\prime}\left( R_{\scr B} \right)} }.
\end{equation*}

\medskip

This barrier has two convenient properties. The first is that it is asymmetric. Like the barriers of the total potential, it decreases
rapidly on the left ($r< R_{\scr B}$) and slowly on the right ($r > R_{\scr B}$).

\medskip

The second important property is that the transmission coefficient through a Morse barrier is known analytically. It is given by 
the expression~\cite{Ahm91}
\begin{equation}
T^{\scr M}(E) = \frac{1 - \exp\left( -4\pi\alpha \right)}{1 + \exp\left[ 2\pi \left(\beta-\alpha\right)\right]},
\label{T0-Morse-exact}
\end{equation}
with
\begin{equation}
\alpha =  ka_{\scr M} = \frac{\sqrt{2\mu E}}{\hbar}\ a_{\scr M}\ \ \ {\rm and}\ \ \ \beta = \frac{\sqrt{2\mu V_{\scr B}}}{\hbar}\ a_{\scr M}.
\label{alpha-beta}
\end{equation}

\subsubsection{Exact barriers vs. approximate barriers}\label{exact vs. approx}

The values of the parameters that fit the Coulomb barriers of the $^4$He + $^{16}$O, $^{12}$C + $^{16}$O and 
$^{16}$O + $^{208}$Pb systems by parabolae and Morse functions are given in table~\ref{tab-barpar}. 
\begin{table}
\caption{
Parameters of the parabolic and of the Morse barriers that best fit the Coulomb barriers of the systems studied in the
present paper. Note that the radius and the height of the barriers are the same for the two parametrizations. They
differ only in the barrier curvature, given by $\hbar\omega$ and by $a_{\scr M}$ in the cases of the parabola and 
the Morse function, respectively.
}
\centering
\begin{tabular} [c] {lccc}
%\begin{tabular} [c] {llll}
\hline \\
%& & \\ 
System:   &   $^4$He + $^{16}$O\ \ \ \ \ \ &   $^{12}$C + $^{16}$O\ \ \ \    & $^{16}$O + $^{208}$Pb   \\ 
 \hline \\
    $V_{\scr B}$ (MeV) $\qquad$ \ \ & 2.90  & 7.99 & 76.55  \\
    $R_{\scr B}$ (fm)     \ \ \ \    \ & 7.42  & 8.02 & 11.59  \\
    $\hbar\omega$ (fm)      \ \ \ \   \ \ & 2.94  & 2.95 & 4.51  \\
     $a_{\scr M}$ (fm)      \ \ \ \   \ \ & 2.96  & 3.13 & 4.32  \\
    \hline 
\end{tabular}
\label{tab-barpar}
\end{table}

\medskip

\begin{figure}%[tb]
\begin{center}
\includegraphics*[width=7 cm]{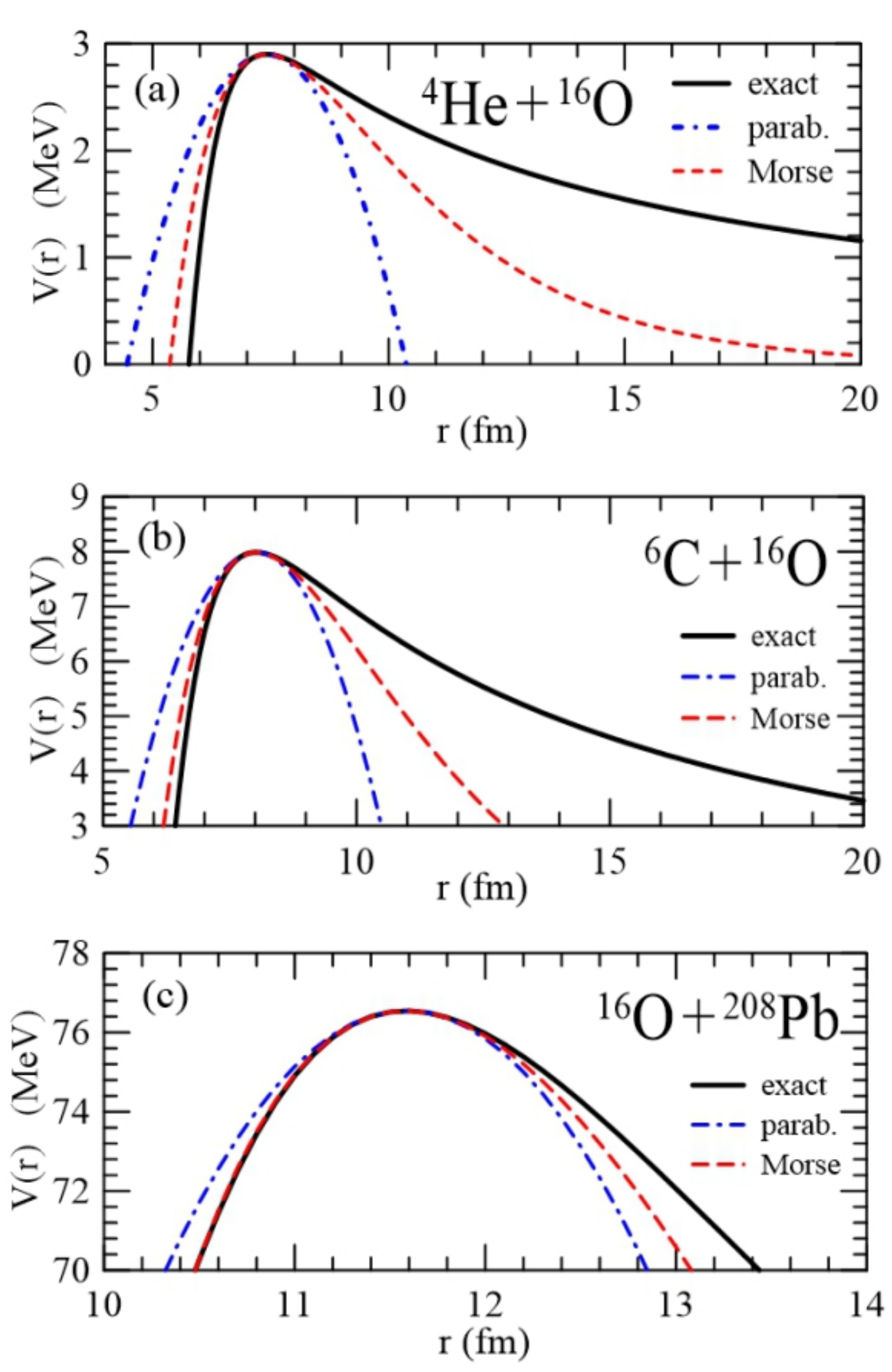}
\end{center}
\caption{(Color on line) Fits of the Coulomb barrier (black solid line) by a parabola (blue dot-dashed line) a Morse function 
(red dashed line). For details see the text.
}
\label{v_parab-morse}
\end{figure}

Fig.~\ref{v_parab-morse} shows the Coulomb barriers (black solid lines) and the best fits by parabolae (blue dot-dashed lines)
and by Morse functions (red dashed lines), for the three systems considered in this paper. The plots show the radial distances 
that influences the transmission coefficients at near-barrier energies.
The figure leads to two conclusions. The first is that the fit by a Morse function is systematically better than by a parabola. This 
is not surprising, since the Morse function is asymmetric, as the Coulomb barrier itself, whereas the parabola is symmetric.  
The second conclusion is that the fits by the two functions are reasonable for $^{16}$O + $^{208}$Pb, but they are poor for
the $^4$He + $^{16}$O and  $^{12}$C + $^{16}$O systems. They become progressively worse as the system's mass decreases. 
The fits are particularly bad at $r \gg R_{\scr B}$, where the Coulomb potential decreases very slowly. Although the Morse function
falls off more slowly on the external side of the barrier, it decreases exponentially, which is much faster than the $1/r$ decay of 
the Coulomb potential. As it will be shown in section~\ref{sect-systems}, the poor fits at large radial
distances lead to dramatic overestimations of transmission coefficients and fusion cross sections at sub-barrier energies. 
%

%%%%%%%%%%%%%%%%%%%%%%%%%%%%%
\subsection{WKB transmission coefficients}
%%%%%%%%%%%%%%%%%%%%%%%%%%%%%

The WKB approximation is a short wavelength limit of Quantum Mechanics. Since it is extensively 
discussed in text books on Quantum Mechanics and Scattering Theory~\cite{Mer98,Joa83,CaH13}, its
derivation will not be presented here. We consider only the application of this approximation in calculations
of transmission coeficients and fusion cross sections.\\ 

Let us consider a nucleus-nucleus collision at a sub-barrier energy $E$, with  angular momentum $\hbar l$. Within the 
WKB approximation, the transmission coefficient is given by
\begin{equation}
T_l^{\scr WKB} (E)= \exp\left[ -2\,\Phi^{\scr WKB}(E) \right],
\label{T0WKB}
\end{equation}
where $\Phi^{\scr WKB}$ is the integral 
\begin{equation}
\Phi^{\scr WKB}(E)  = \int_{r_1}^{r_2} \kappa(r)\ dr,
 \label{PhiWKB}
\end{equation}
with
\begin{equation}
\kappa(r) = \frac{\sqrt{2\mu\,\big[ V_l(r)-E\big] }}{\hbar}.
\label{kappa}
\end{equation}
In the above equations $\mu$ is the reduced mass of the projectile-target system, $V_l(r)$ is the total potential of 
Eq.~(\ref{Veff}), and $r_1$ and $r_2$ are the classical turning points. They are the solutions of the equation on $r$,
\begin{equation}
V_l(r) =E.
\label{turningpoints}
\end{equation}
The influence of multiple reflections under the barrier, ignored in the above equations, was investigated by Brink and 
Smilansky~\cite{BrS83}. They concluded that these reflections are relevant at energies just below the barrier, contributing 
to improve the agreement with the exact results. \\

Although Eq.~(\ref{T0WKB}) is a very good approximation to the exact  transmission coefficient at energies well below the 
barrier of the total potential, denoted by $B_l$, it becomes progressively worse as the energy approaches $B_l$.
Furthermore, at energies above the barrier, the WKB transmission coefficient takes the constant value $T_l^{\scr WKB}= 1$,  
whereas its quantum mechanical counterpart is equal to 1/2 at the barrier, and grows continually to one as the energy increases. 

\bigskip

In 1935, Kemble~\cite{Kem35} showed that the WKB approximation can be improved if one uses a better connection formula. 
He got the expression
\begin{equation}
T_l^{\scr K} (E)= \frac{1}{1+\exp\left[2\,\Phi^{\scr WKB}(E) \right]}.
\label{T0Kemble}
\end{equation}
At energies well below the Coulomb barrier, the two turning points are far apart and $\kappa(r)$ reaches appreciable values 
within the integration limits of Eq.~(\ref{PhiWKB}). In this way, $\Phi^{\scr WKB}(E)$ becomes very large, so that the unity 
can be neglected in the denominator of Eq.~(\ref{T0Kemble}). This equation then reduces to Eq.~(\ref{T0WKB}). Therefore,
the two approximations are equivalent in this energy region. However, Kemble's approximation remains valid as the energy
approaches the barrier, leading to the correct result at $E=B_l$, namely $T_l=1/2$.

%%%%%%%%%%%%%%%%%%%%%%%%%%%%%%%%%%%%%%%
\subsection{Kemble transmission coefficient at above-barrier energies}
%%%%%%%%%%%%%%%%%%%%%%%%%%%%%%%%%%%%%%%

The problem with  WKB approximations (both  standard and Kemble's version) at above-barrier energies
is that there are no classical turning points. At $E=V_{\scr B}$ the two turning points coalesce, and above this limit
Eq.~(\ref{turningpoints}) has no real solution. Then, $ \Phi^{\scr WKB}\left( E>V_{\scr B} \right)= 0$, and the transmission
coefficients of Eqs.~(\ref{T0WKB}) and (\ref{T0Kemble}) take respectively the constant values $T_{\scr l}^{\scr WKB} = 1$ 
and $T_{\scr l}^{\scr K} = 1/2$. \\

However, Kemble~\cite{Kem35} pointed out (see also the book by Fr\"oman and Fr\"oman~\cite{FrF65}, where 
this problem is treated formally) that Eq.~(\ref{T0Kemble}) can be extended to above-barrier energies if one solves 
Eq.~(\ref{turningpoints}) in the complex r-plane, and evaluates the integral between the complex turning points. 
Although Kemble ~\cite{Kem35} did not discuss this analytical continuation in detail, he pointed out that it would
lead to the exact expression for the parabolic barrier below and above the barrier. More recently, the analytical
continuation in the case of a typical heavy ion potential was carried out numerically~\cite{TCH17}, and the resulting
transmission coefficient was shown to be in very good agreement with its quantum mechanical counterpart.
In the next section we carry out this analytical continuation in the cases of the parabolic and
the morse barriers, where all calculations can be performed analytically. 

%%%%%%%%%%%%%%%%%%%%%%%%%%%%%%%%%%%%%%%
\subsubsection{Transmission through a parabolic barrier for $E>V_{\scr B}$}
%%%%%%%%%%%%%%%%%%%%%%%%%%%%%%%%%%%%%%%

Since the transmission coefficient through a parabolic barrier is known exactly, it is an ideal test 
for the analytical continuation procedure.
However, before any consideration involving the complex plane, we prove that Kemble approximation for a 
parabolic barrier is exact at sub-barrier energies. For simplicity, we discuss in the section the particular case
of a S-wave. The extension to other values of the angular momentum is straightforward. One has just to use
the parameters of the $l$-dependent barrier, replacing: $V_{\scr B}\rightarrow B_l$, $R_{\scr B} \rightarrow R_l$
and $\hbar\omega \rightarrow \hbar\omega_l$.\\

Since the exact transmission coefficient of Eq.~(\ref{T0HW}) 
has the same general form of the Kemble transmission coefficient (Eq.~(\ref{T0Kemble})), the two expressions 
will be identical if 
\begin{equation}
\Phi^{\scr WKB}(E) = \Phi^{\scr HW}(E).
\label{equiv}
\end{equation}
Using the explicit forms of $\Phi^{\scr WKB}$ (Eq.~(\ref{PhiWKB})) and $\Phi^{\scr HW}$ (Eq.~(\ref{PhiHW})), the above 
equation becomes,
\begin{multline}
 \frac{\sqrt{2\mu}}{\hbar}  \int_{r_-}^{r_+} dr\ \sqrt{ 
V_{\scr B} - E- 
\frac{\mu \omega^2}{2} \left(r-R_{\scr B}\right)^2}   \\
                          = \frac{\pi}{\hbar\omega}\ \left(V_{\scr B} -E \right).
\label{rel1-a}
\end{multline}
To calculate this integral we make the transformation: 
\begin{equation*}
r\  \longrightarrow\           x = \omega\,\sqrt{\frac{\mu}{2}}
%\frac{\sqrt{2\mu}}{\hbar}
\ \left(r-R_{\scr B} \right),
%\label{transf_r-x}
\end{equation*}
so that
\[
dr = \frac{1}{\omega}\,\sqrt{\frac{2}{\mu}}\ \ dx, \ \  r_\pm \rightarrow x_\pm = \pm \sqrt{V_{\scr B} -E}. 
\]
Eq.~(\ref{rel1-a}) then becomes
\begin{equation}
\Phi^{\scr WKB} =   \frac{2}{\hbar\omega}\ \int_{x_-}^{x_+} dx\ \sqrt{ \left(V_{\scr B} - E \right) - x^2} .
\label{ref2}
\end{equation}
This integral can be easily evaluated and the result is
\begin{equation}
\Phi^{\scr WKB} = \frac{\pi}{\hbar\omega}\ \left( V_{\scr B} - E \right).
\label{ref3}
\end{equation}
Thus,  we conclude that Kemble approximation for a parabolic barrier is exact at sub-barrier energies.\\

Now we consider collisions at above-barrier energies. To deal with this situation, we carry out the analytical continuation 
of $x$ to the complex plane. Setting $x \rightarrow z =x+i\,y$, the parabolic barrier becomes,
\begin{equation}
V(z) = V_{\scr B} -  z^2.
\label{z-parab}
\end{equation}
The turning points are then the complex solutions of the equation,
\begin{equation}
V(z) - E = 0 ,
\label{complextp}
\end{equation}
for $E>V_{\scr B}$. Clearly, these turning points must be located on the complex r-plane, in a region where the potential 
is real. Solving Eq.~(\ref{complextp}) for a general potential is not simple. However, it can be easily done for a 
parabolic barrier, as that of Eq.~(\ref{z-parab}). In this case, $V(z)$ is real only on the $x$ and on the $y$ axes. 
Therefore, the turning points must be either real or imaginary. The real turning points are the solutions of 
Eq.~(\ref{complextp}) for $E<V_{\scr B}$, which we have already discussed. Now we consider the imaginary
solutions. Inserting Eq.~(\ref{z-parab}) into Eq.~(\ref{complextp}) and setting $z= iy$, one gets the coordinates of the 
two imaginary turning points,
\begin{equation*}
y_{\scr \pm} = \pm\, \sqrt{E-V_{\scr B}}.
\end{equation*}
\begin{figure}%[tb]
\begin{center}
\includegraphics*[width=6cm]{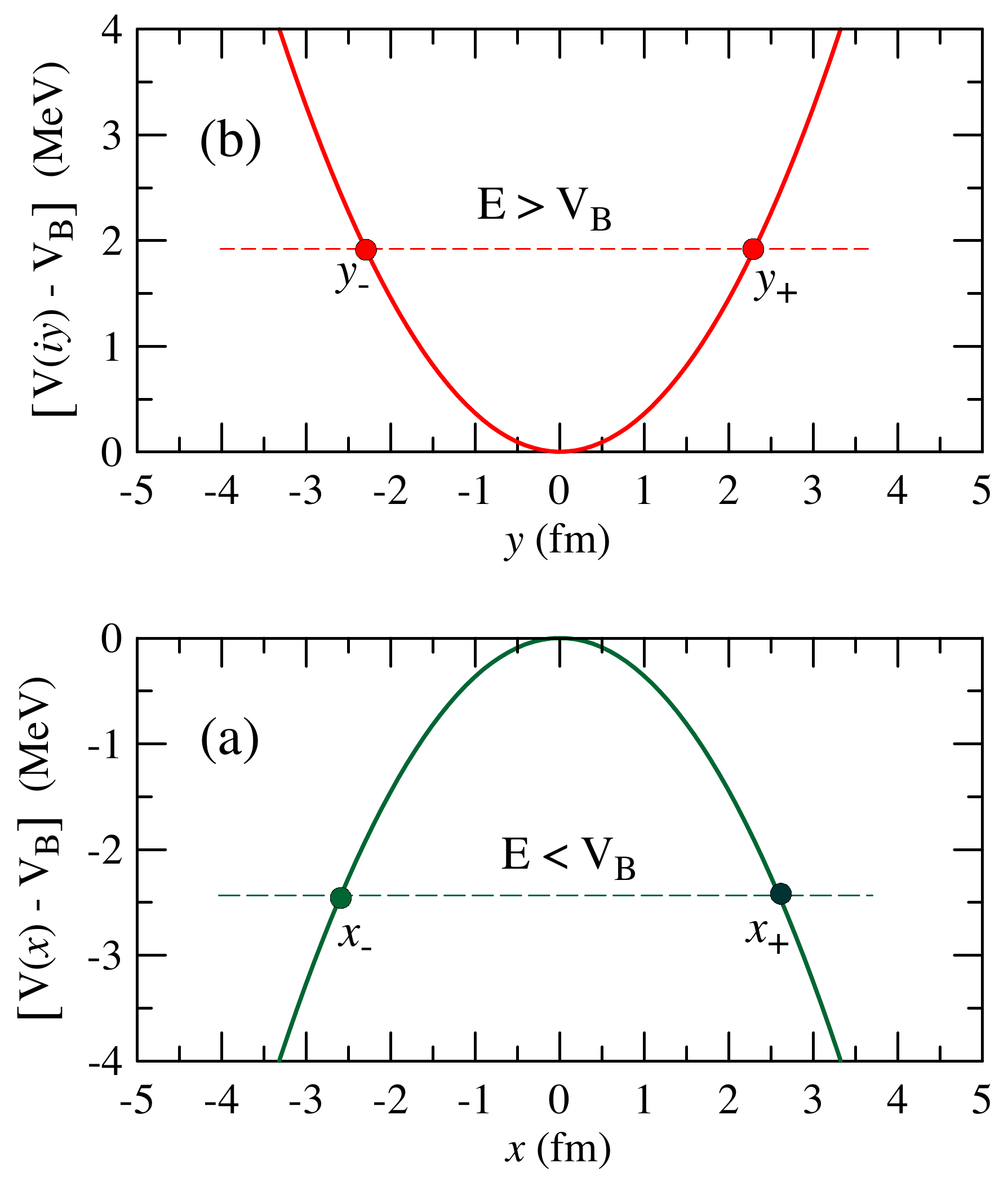}
\end{center}
\caption{(Color on line) Analytical continuation of the parabolic potential barrier $V(x+iy)$. Panels (a) and (b) show respectively the potential on the real 
($y=0$) and on the imaginary ($x=0$) axes. The figure also show the real and the imaginary turning points in the cases of $E<V_{\scr B}$ (panel (a)) and
$E>V_{\scr B}$ (panel (b)). }
\label{par-invpar}
\end{figure}
Fig.~\ref{par-invpar} shows a parabolic potential barrier (panel (a)) and its analytical continuation (panel (b)). The green
circles on panel (a), $x_-$ and $x_+$, represent the real turning points for an energy $E<V_{\scr B}$. The red circles in 
panel (b)), $y_-$ and $y_+$, indicate the imaginary turning points for an energy $E>V_{\scr B}$.

\bigskip

With the analytical continuation discussed above, the WKB integral of Eq.~(\ref{ref2}) can be extended to the complex plane. It 
must be evaluated along the imaginary axis, between the turning points $z_{\scr \pm} = iy_{\scr \pm}$. In this case, the integrand
of Eq.~(\ref{PhiWKB}) must be generalized as
\begin{equation}
\Phi^{\scr WKB}  = \int_{r_-}^{r_+} \kappa (r)\,dr\ \rightarrow\ \Phi^{\scr WKB}  =  \int_{z_-}^{z_+}
 \kappa(z)\,dz  .
\label{Re(kappa dy)}
\end{equation}
Using the explicit form of the potential (Eq.~(\ref{z-parab})) in Eq.~(\ref{kappa}) for $z=iy$, and changing the integration variable to $y$, one gets,
\begin{equation*}
\Phi^{\scr WKB} =   - \frac{2}{\hbar\omega}\ \int_{y_-}^{y_+} \,dy\ \sqrt{ \left(E-V_{\scr B} \right) - y^2}                          .
%\label{ref4}
\end{equation*}
This integral is equivalent to the one in Eq.~(\ref{ref2}), and the result is the same. Thus, we have shown
that Kemble's formula for the transmission coefficient through a parabolic barrier is valid for any collision energy. Besides,
it gives the exact quantum mechanical result.

\subsubsection{Transmission through a Morse barrier for $E>V_{\scr B}$}
 
 As in the previous section, we present a detailed discussion of the S-wave transmission coefficient. To extend it to 
 other angular momenta, one has just to use the Morse parameter of the $l$-dependent barrier, changing:
  $V_{\scr B}\rightarrow B_l$,  $R_{\scr B}\rightarrow R_l$ and $a_{\scr M}\rightarrow a_l$.\\
 
We start with an important remark about the Morse approximation for the Coulomb barrier. In typical heavy 
ion collisions at near-barrier energies, the following relation is satisfied:
\begin{equation*}
f(\mu,E) \equiv \exp\left( - 4\pi\,\alpha \right) \equiv \exp\left( 
-\frac{4\pi\,a_{\scr M}\ \sqrt{2\mu E}}{\hbar} 
\right) \ll 1.
\end{equation*} 
This term falls exponentially with the factor $\sqrt{\mu E}$. Thus, its largest values are for the lightest system, at 
the lowest collision energy. In the present study it corresponds to $^4$He + $^{16}$O, at 0.5 MeV. Under these
conditions one gets, $f(\mu,E) = 3\times 10^{-5}$. For the other two systems, this term is several orders of magnitude
smaller. Therefore, it can be safely neglected, and Eq.~(\ref{T0-Morse-exact}) reduces to
\begin{equation}
T^{\scr M}(E) = \frac{1}{1 + \exp\left[ 2\Phi^{\scr M}(E)\right]}.
\label{T0-Morse-1}
\end{equation}
Above,
\begin{equation}
\Phi^{\scr M}(E) = \pi \left(\beta-\alpha\right),
\label{Phi-Morse}
\end{equation}
where $\alpha$ and $\beta$ are given by Eq.~(\ref{alpha-beta}).
Comparing Eqs.~(\ref{T0-Morse-1}) and Eq.~(\ref{T0Kemble}), one concludes that Kemble's approximation for the Morse
barrier will be exact if the following condition is satisfied:
\begin{equation}
\Phi^{\scr M}(E)=\Phi^{\scr WKB}(E).
\label{M-WKB}
\end{equation}
Using Eqs.~(\ref{PhiWKB}) and (\ref{Phi-Morse}), the above condition becomes,  
\begin{equation}
 \frac{\sqrt{2\mu}}{\hbar}\ \int_{r_-}^{r_+} dr\  \sqrt{V_{\scr M}(r)  -E } = \pi \left(\beta-\alpha\right),
\label{T0-Morse-2}
\end{equation}
where $r_\pm$ represent the solutions of the equation
\begin{equation*}
V_{\scr M}\left(r_\pm \right) - E=0.
\end{equation*}
These solutions can easily be determined and one finds,
\begin{equation*}
r_{\pm} = R_{\scr B}-a_{\scr M}\, \ln\left[ 1\pm \sqrt{1-\varepsilon }\right],
\end{equation*}
where we have introduced the notation
\begin{equation*}
\varepsilon = \frac{E}{V_{\scr B} }.
\end{equation*}
To check the validity of Eq.~(\ref{M-WKB}), we evaluate the WKB integral of Eq.~(\ref{T0-Morse-2}). Using the explicit form 
of the Morse potential (Eq.~(\ref{Vmorse})) and changing to the new variable
\begin{equation*}
r\ \longrightarrow\ t=\frac{\exp\big[- (r-R_{\scr B})/a_{\scr M}\big] - 1}{\sqrt{1-\varepsilon}} ,
\end{equation*}
the integral takes the form
\begin{equation*}
\Phi^{\scr WKB}(E)  = \frac{\sqrt{2\mu V_{\scr B}}}{\hbar}\ a_{\scr M}\ \left(1-\varepsilon\right)\ \int_{-1}^{1} dt \  \frac{
\sqrt{1-t^2 }} {1+t\,\sqrt{1-\varepsilon}}.
%\label{T0-Morse-3}
\end{equation*}
This integral can be evaluated analytically and the result is
\begin{equation}
\Phi^{\scr WKB}(E)   = \pi \left(\beta-\alpha\right),
\label{Phi-Morse-2}
\end{equation}
which coincides with the expression for $\Phi^{\scr M}(E)$ (Eq.~(\ref{Phi-Morse})).\\

However, the above proof is not valid for $E>V_{\scr B}$. The reason is that $\Phi^{\scr WKB}$  was evaluated by an
integration between two real turning points (see Eq.~(\ref{T0-Morse-2})), which do not exist in this energy region. Nevertheless, the
validity of Eq.~(\ref{T0-Morse-2}) can be extended to $E >V_{\scr B}$, through an analytical continuation of the variable $r$. This 
procedure, which is analogous to the one adopted for a parabolic barrier, will be followed below. \\

First, we introduce the new projectile-target distance variable,
\begin{equation*}
x = \frac{r-R_{\scr B}}{a_{\scr M}},
\end{equation*}
where the Morse parameter, $a_{\scr M}$, and the barrier radius, $R_{\scr B}$, are known quantities. Then, we perform the analytical continuation
of $x$ to the complex plane. That is, $x \rightarrow z=x+i\,y$. Using the explicit expression of the potential in terms 
of $x$ and $y$, one gets
\begin{eqnarray*}
V(z) &=& V_{\scr B}\ \left[ 2\,\exp\left(-x-iy \right) - \exp\left(-2x-2iy \right)\right]\nonumber\\
       &=& U(x,y)+i\ W(x,y),
\end{eqnarray*}
with
\begin{equation}
U(x,y) = V_{\scr B}\, \Big[ 2\,e^{-x}\,\cos y\ \left(1- e^{-x}\,\cos y\right) + e^{-2x}  \Big]
\label{Umorse}
\end{equation}
and
\begin{equation}
W(x,y) = 2\,V_{\scr B}\, e^{-x} \sin y\ \big[ e^{-x}\,\cos y-1  \big] .
\label{Wmorse}
\end{equation}
Since in the equation defining the turning points the potential must be real, we set 
\begin{equation}
W(x,y)=0.
\label{W=0-1}
\end{equation}
The solutions of the above equation are
\begin{equation}
\sin y =0
\label{cond1}
\end{equation}
and
\begin{equation}
e^{x} = \cos y .
\label{cond2}
\end{equation}
Eq.~(\ref{cond1}) is satisfied on the real axis and on other horizontal lines intercepting the y-axis at $y=\pm\, n\pi$, where $n$ is any integer.
It can be easily checked that the potential evaluated at any point on these lines cannot be higher than $V_{\scr B}$. Thus, there are no 
turning points on them. Therefore, these solutions must be discarded. \\

\begin{figure}%[tb]
\begin{center}
\includegraphics*[width=8 cm]{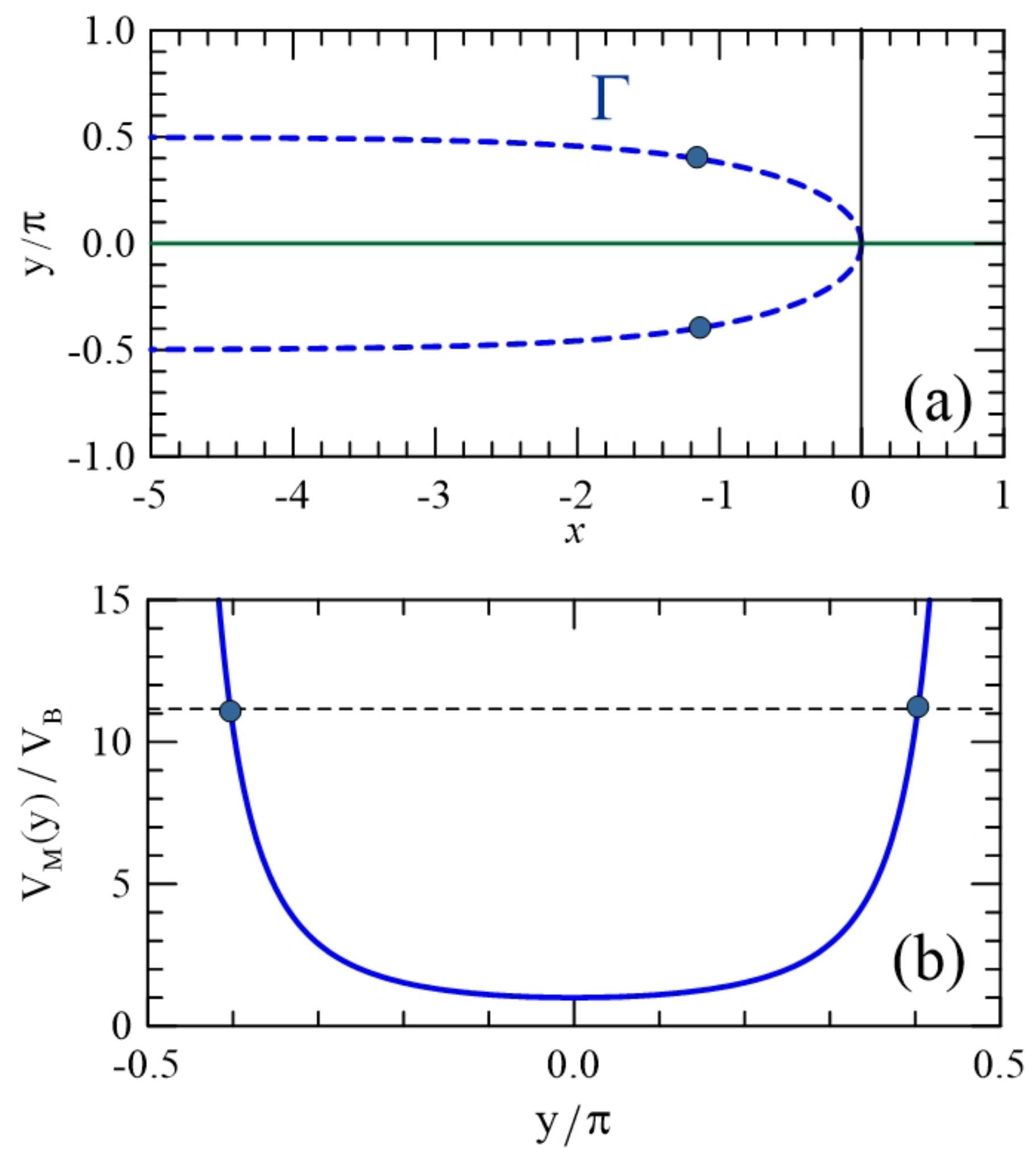}
\end{center}
\caption{(Color on line) The analytic continuation of the Morse potential on the complex plane. Panel (a) shows the lines where the Morse 
potential is real. The green solid line is the trivial solution of Eq.~(\ref{cond1}) (the real-axis), whereas the blue dashed line, labelled by 
$\Gamma$, is the solution of Eq~(\ref{cond2}). The solid circles represent the turning points for an arbitrary energy above the barrier. Panel
(b) shows the potential for points on $\Gamma$, where it is real, divided by $V_{\scr B}$.
}
\label{vmorsez-fig}
\end{figure}

We are then left with the solutions of Eq.~(\ref{cond2}). They are curves on the complex plane confined to the left half-plane ($x<0$). 
Similarly to Eq.~(\ref{cond1}), the periodicity of the trigonometric function (here $\cos y$) leads to an infinite number of solutions. They are
curves that can be obtained from one another by shifts of $2\pi$ along the y-axis. Nevertheless, they lead to the same physics. Therefore we 
concentrate on the one corresponding to the lowest values of $|y|$. This curve, denoted by $\Gamma$,  is represented on panel  (a) of 
Fig.~(\ref{vmorsez-fig}). The turning points for $E=2\,V_{\scr B}$ are represented by solid circles.

\bigskip

On the curve $\Gamma$, the variables $x$ and $y$ are not independent. They are related by Eq.~(\ref{cond2}). Thus, the potential becomes a function 
of a single variable. The coordinate $x$ along this curve is the single-valued function of $y$, 
\begin{equation}
x_{\scr \Gamma}(y) = \ln\left(\cos y \right).
\label{x_Gamma}
\end{equation}
Owing to the infinite values of the potential at $y=\pm \pi/2$, the coordinate $y$ is confined to the open interval $\left(-\pi/2,\pi/2 \right)$. Then, 
$\cos y$ is positive, so that the solution of Eq.~(\ref{x_Gamma}) is well defined. 

\medskip

To obtain the real potential for points on $\Gamma$, $U\left( x_{\scr \Gamma},y \right)$, one inserts Eq.~(\ref{x_Gamma}) into 
Eq.~(\ref{Umorse}). One gets, \\
\begin{equation}
U\left( x_{\scr \Gamma},y \right) \equiv U_{\scr \Gamma}(y)=V_{\scr B}\ \Big[1+\tan^2 y \Big].
\label{Uxy}
\end{equation}

Now we evaluate the WKB integral on the complex plane,
\begin{equation}
\Phi^{\scr WKB}(E) =a_{\scr M}\ \frac{\sqrt{2\mu}}{\hbar}\  \int_{z_{\scr -}}^{z_{\scr +}} 
dz\ \sqrt{V_{\scr M} \left( z_{\scr \Gamma} \right) - E }\,.
\label{int-z}
\end{equation}
We remark that the factor $a$ results from the change of variable $r\rightarrow z=(r-R_{\scr B})/a_{\scr M}$. For practical purposes, it is 
convenient to evaluate the integral over the contour $\Gamma$. On this contour, $V_{\scr M}(x,y)$ reduces to the real potential 
of Eq.~(\ref{Uxy}), $U_{\scr \Gamma}(y)$,  and the differential $dz$ can be written as
\begin{equation*}
dz = \left[\frac{dx_{\scr \Gamma}(y)}{dy} + i\right]\ dy,
\end{equation*}
where $dx_{\scr \Gamma}(y)/dy$ is a real function of $y$. 
For energies above the barrier, $U_{\scr \Gamma}(y) - E$ is negative so that,
\begin{eqnarray}
dz\ \sqrt{U_{\scr \Gamma}(y) -E} &=& \left( \frac{dx_{\scr \Gamma}(y)}{dy} + i\right) \times i\ \sqrt{E-U_{\scr \Gamma}(y)} \ dy \nonumber\\
                                                    &=&  G_{\rm R}(y) \ dy + i\, G_{\rm I}(y)\ dy,
                                                    \label{GR-GI}
\end{eqnarray}
where $G_{\rm R}(y)$ and  $G_{\rm I}(y)$ are the real functions,
\begin{eqnarray}
G_{\rm R}(y) &=&  - \sqrt{E-U_{\scr \Gamma}(y)} \label{GR}, \\
G_{\rm I}(y) &=&  \frac{x_{\scr \Gamma}(y)}{dy}\ \, \sqrt{E-U_{\scr \Gamma}(y)} .
\label{GI}
\end{eqnarray}
The values of $y$ corresponding to the integration limits of Eq.~(\ref{int-z}) are given by the equation,
\begin{equation*}
U_{\scr \Gamma}(y) = E,\ \ {\rm or}\ \ 1+\tan^2 y = \varepsilon ,
\end{equation*}
which has the solutions,
\begin{equation*}
y_\pm =\pm \tan^{-1}\sqrt{1-\varepsilon} .
\label{y_pm}
\end{equation*}

Expressing the integral of Eq.~(\ref{int-z}) in terms of the variable $y$, one gets
\begin{multline}
\Phi^{\scr WKB}(E) =a_{\scr M}\, \frac{\sqrt{2\mu}}{\hbar}\   \int_{y_{\scr -}}^{y_{\scr +}} G_{\rm R}(y)\ dy\, \\
+\, i \int_{y_{\scr -}}^{y_{\scr +}} G_{\rm I}(y)\ dy \,.
\label{GR-GI}
\end{multline}
Inspecting Fig.~\ref{vmorsez-fig}, one concludes that $G_{\rm R}(y)$ is an even function of $y$, whereas $G_{\rm I}(y)$ is an odd
function of $y$. Since the integration limits are symmetrical, the integration of $G_{\rm I}(y)$ vanishes. Then, using the explicit form of 
$G_{\rm R}(y)$ (Eq.~(\ref{GR})) with the potential of Eq.~(\ref{Uxy}), one gets
\begin{equation}
\Phi^{\scr WKB}(E) = - a_{\scr M}\ \frac{\sqrt{2\mu\,V_{\scr B}}}{\hbar}\  \int_{y_{\scr -}}^{y_{\scr +}} dy\ 
\sqrt{(\varepsilon-1)\,-\, \tan^2 y} \ .
\label{int-y-0}
\end{equation}
To evaluate this integral, we change to the new variable, $y \rightarrow t = \tan y$, so the above integral becomes,
\begin{equation}
\Phi^{\scr WKB}(E) = - a_{\scr M}\ \frac{\sqrt{2\mu\,V_{\scr B}}}{\hbar}\ 
 \int_{-\sqrt{\varepsilon -1}}^{\sqrt{\varepsilon -1}}\  dt\ 
\frac{
\sqrt{(\varepsilon-1) - t^2} } {1+t^2}
\ .
\label{int-t}
\end{equation}
The above integral can be found in standard integral tables and the result is \\
\begin{eqnarray*}
\Phi^{\scr WKB}(E) & = & \pi\ \frac{a_{\scr M}\, \sqrt{2\mu V_{\scr B}}}{\hbar}\ \big(1-\sqrt{\varepsilon} \big)\nonumber\\
                               & = & \pi\ \ \big( \beta - \alpha \big).
%\label{PhiWKB-M} 
\end{eqnarray*}

Thus, we have proved that Kemble's transmission coefficient through a Morse barrier reproduces  accurately the exact quantum mechanical 
result, both below and above the Coulomb barrier.  \\

%%%%%%%%%%%%%%%%%%%%%%%%%%%%%%%%%%%%%%%%%%%%%%%%%%%%%%%%%%%%%%%%%%%%%%

\subsection{Wong's approximation for the fusion cross section}\label{Wong-sect}

%%%%%%%%%%%%%%%%%%%%%%%%%%%%%%%%%%%%%%%%%%%%%%%%%%%%%%%%%%%%%%%%%%%%%%

In 1973 Wong~\cite{Won73} proposed a simple expression for the fusion cross section, in which the Coulomb barrier is 
approximated by a parabola, and the angular momentum is treated as a classical variable. Wong's formula is based
on the following assumptions:

\begin{enumerate}

\item The fusion probability at the $l^{\rm th}$ partial- wave is approximated by the Hill-Wheeler transmission factor
\begin{equation*}
P_l^{\scr F}(E) \simeq  \frac{1}{1+\exp\left[ 2\pi \left( B_l- E  \right)/\hbar\omega_l\right]},
%\label{PFl-Wong}
\end{equation*}
with $B_l$ and $\hbar\omega_l$ standing for the height and the curvature parameters of the parabolic approximation for the
barrier of $V_l(r)$.

\item The radii and the curvature parameters of the $l$-dependent barriers where assumed to be independent of $l$. That is
$R_l = R_{l=0} \equiv R_{\scr B}$ and $\hbar\omega_l = \hbar\omega_{l=0} \equiv \hbar\omega$. With this assumptions,
the barrier height takes the simple form,
\begin{equation}
B_l = V_{\scr B}+ \frac{\hbar^2\, l(l+1)}{2\mu\,R_{\scr B}^2}.
\label{Bl}
\end{equation}.

\item The angular momentum is treated as the continuous variable $l \rightarrow \lambda =l+1/2$. Then, one approximates:
$l(l+1) \simeq \lambda^2$ and $\sum_l (2l+1)\rightarrow 2\,\int\ d\lambda\  \lambda$.
\end{enumerate}

\bigskip

With these simplifying assumptions, Eq.~(\ref{lsumsigF}) becomes
\begin{equation*}
\sigma_{\scr F}(E) =   \frac{1}{E}\ \frac{\pi\hbar^2}{\mu}\ \int_0^\infty d\lambda\  \lambda\ T(\lambda,E),
\end{equation*}
where
\begin{equation*}
T(\lambda,E) =  \frac{1}{1+\exp\left[ \frac{2\pi}{\hbar\omega_l} \left( V_{\scr B}- E + \frac{\hbar^2\lambda^2}{2\mu R_{\scr B}^2 } 
 \right) \right]}.
\end{equation*}
The above integral can be evaluated analytically and the result is Wong's cross section, which can be written as 
\begin{equation}
\sigma_{\scr F}^{\scr W}(E) =  R^2_{\scr B}\ \frac{\hbar\omega}{2 E}\ F_0(x).
\label{wong1-a}
\end{equation}
Above, $x$ is the modified energy variable
\begin{equation}
x = \frac{E-V_{\scr B}}{\hbar\omega},
\label{wong1-b}
\end{equation}
and $F_0(x)$ is the dimensionless and system independent function
\begin{equation}
F_0(x) =   \ln \left[ 1 + \exp \left(  2\pi\, x \right)  \right].
\label{wong1-c}
\end{equation}
This function is known in the literature as the {\it universal fusion function}. It is frequently used as a benchmark 
in comparative studies of fusion reactions~\cite{CGL09b,CGL09a,CGD15}. \\

\subsubsection{Energy-dependence of the barrier parameters}\label{Edep_W}

Rowley and Hagino~\cite{RoH15} pointed out that the barrier radius of $V_l(r)$ may decrease appreciably with
$l$, mainly for light heavy-ion systems. In this case, the radius for the grazing angular momentum may be a few
fermi smaller that that for $l=0$. Then, the approximation $R_l\simeq R_{\scr B}$ is poor, and this makes Wong's formula
inaccurate. The situation gets worse as the energy increases, so that $l_{\scr E}$ takes large values. To cope with this
situation, Rowley and Hagino proposed an improved version of the Wong formula, where the S-wave barrier parameters 
are replaced by the parameters associated with the grazing angular momentum.

\medskip

The grazing angular momentum at the energy $E$, which we denote by $\lambda_{\scr E}$, and the corresponding barrier radius,
$R_{\scr E}$, are given by the coupled equations:
\begin{equation}
V_{\lambda_{\scr E}}(r) = V_{\scr N}(R_{\scr E}) + V_{\scr C}(R_{\scr E}) + \frac{\hbar^{2}\,\lambda_{\scr E}^2 }
{2\mu R_{\scr E}^2}   = E,
\end{equation}
and 
\begin{equation}
\left[ \frac{d V_{\lambda_{\scr E}}(r)}{dr}\right]_{R_{\scr E}} = 0.
\end{equation}
Solving these equations, one determines $l{\scr E}$ and $R{\scr E}$, and the barrier curvature parameter is given by\\
\begin{equation*}
\hbar\omega_{\scr E} = \sqrt{
\frac{ - \, \hbar^2\,V_{\lambda_{\scr E}}^{\prime\prime}\left( R_{\scr E}\right)} {\mu}}.
%\label{hbaromega}
\end{equation*}

The above equations supply the barrier parameters for the grazing angular momentum, which depends on the collision energy. 
 Accordingly, the improved Wong formula becomes,
\begin{equation}
\sigma_{\scr F}^{\scr W}(E) =  R^2_{\scr E}\ \frac{\hbar\omega_{\scr E}}{2 E}\  F_0(x_{\scr E}),
\label{Nsig}
\end{equation}
where,
\begin{equation}
F_0(x_{\scr E}) =   \ln \left[ 1 + \exp \left(  2\pi\, x_{\scr E} \right)  \right].
\label{NF}
\end{equation}
and,
\begin{equation}
x_{\scr E} = \frac{E - V_{\scr E}}{\hbar\omega_{\scr E}}.
\label{Nx}
\end{equation}
Above, we have introduced the `effective Coulomb barrier' for the energy $E$,
\begin{equation}
V_{\scr E} = B_\lambda -  \frac{\hbar^{2}\,\lambda_{\scr E}^2 }
{2\mu R_{\scr E}^2} = V_{\scr N}(R_{\scr E}) + V_{\scr C}(R_{\scr E}).
\label{VE}
\end{equation}
The above expression is valid for $B_{\lambda_{\scr crit}} \ge E \ge V_{\scr B}$, where $\lambda_{\scr crit}$ is the
critical angular momentum, corresponding to the largest angular momentum where the total potential has a pocket. For energies 
below $V_{\scr B}$, one sets $V_{\scr E} = V_{\scr B},\, R_{\scr E} = R_{\scr B}$ and $\omega_{\scr E}=\omega$. For energies 
above $B_{\lambda_{\scr crit}}$,  $V_{\scr E}$ is given by Eq.~(\ref{VE}) but with $B_\lambda$ replaced by $B_{\lambda_{\scr crit}}$,
and one sets $R_{\scr E} = R_{\lambda_{\scr crit}}$ and $\omega_{\scr E}=\omega_{\lambda_{\scr crit}}$.

%%%%%%%%%%%%%%%%%%%%%%%%%%%%%%%%%%%%%%%%%%%%%%%%%%%%%%%%%%%%%%%%%
\section{Applications} \label{sect-systems}
%%%%%%%%%%%%%%%%%%%%%%%%%%%%%%%%%%%%%%%%%%%%%%%%%%%%%%%%%%%%%%%%%

Now we discuss the use of the approximations of the previous sections to evaluate S-wave transmission coefficients and fusion
cross sections. We perform calculations for a very light system and two slightly heavier ones, namely $^6$He + $^{16}$O, 
$^{12}$C + $^{16}$O and $^{16}$O + $^{208}$Pb.

%%%%%%%%%%%%%%%%%%%%%%%%%%%%%%%%
\subsection{S-wave transmission coefficients} 
%%%%%%%%%%%%%%%%%%%%%%%%%%%%%%%%

%
\begin{figure}%[tb]
\begin{center}
\includegraphics*[width=6.5 cm]{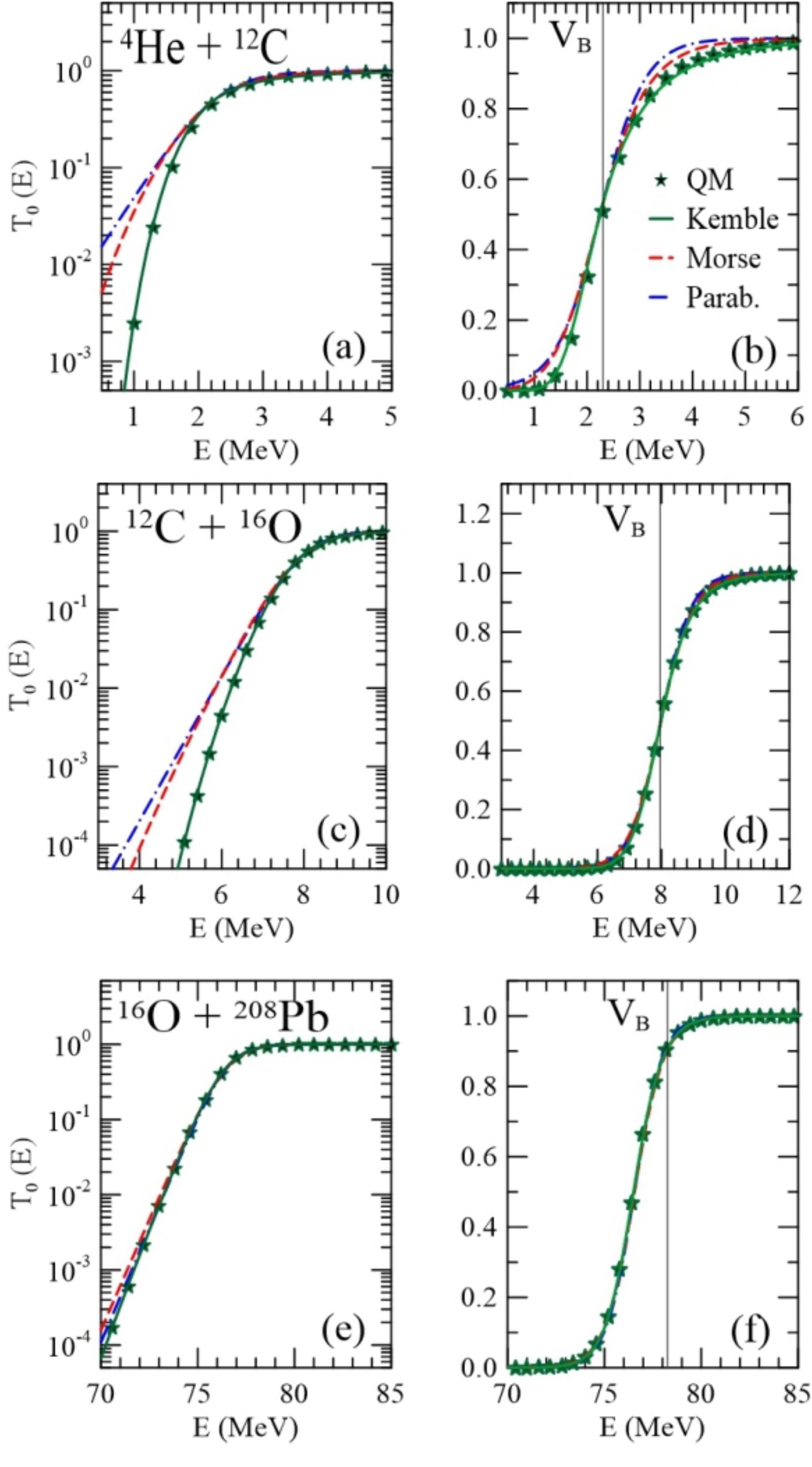}
\end{center}
\caption{(Color on line) S-wave transmission coefficients for the $^4$He + $^{16}$O, $^{12}$C + $^{16}$O and $^{16}$O + $^{208}$Pb at
near-barrier energies. The figure shows the results for a parabolic barrier (blue dot-ashed lines), for a Morse barrier (red dashed lines) and
the predictions of Kemble's WKB approximation for the actual barrier (green solid lines), in comparison with the quantum mechanical results
(stars). The results are shown in logarithmic (a,c,e) and linear (b,d,f) scales.
}
\label{T0_all}
\end{figure}
Fig.~\ref{T0_all} shows S-wave transmission coefficients obtained with the approximations discussed in the previous sections, in comparison
with the exact quantum mechanical transmission coefficients (stars). The notation for the approximations are indicated in panel (b).  The
results are shown in logarithmic scales (left panels), which is appropriate to compare cross sections at sub-barrier energies, and in linear
scales (right panels), which gives a better picture at energies above the barrier. 
In the calculations of Kemble's transmission coefficients above the barrier, we used the elliptical approximation for the 
curves of real potential on the complex $r$-plane. As shown in Ref.~\cite{TCH17}, this approximation leads to accurate results, while it
simplifies considerably the calculations.  

\medskip

In the case of the heaviest system, $^{16}$O + $^{208}$Pb, Kemble WKB (with the analytical continuation of $r$~\cite{TCH17}) reproduces the quantum
mechanical transmission coefficient with great accuracy, above and below the Coulomb barrier. The other two curves, corresponding to approximations
of the Coulomb barrier by a parabola (blue dot-dashed lines) and by a Morse function (red dashed lines) are very close to each other. They are also 
close to the exact results, except for the lowest energies in the plot, where the two approximations overestimate slightly the quantum mechanical results.

\medskip

The situation is different for the $^4$He + $^{16}$O and $^{12}$C + $^{16}$O systems. Although Kemble's WKB remains very accurate, above and below the 
barrier, the parabolic and the Morse approximations are very poor at sub-barrier energies. These approximations greatly overestimate the transmission
coefficients, by more than one order of magnitude at the lowest energies in the plots. In the case of the lightest system, $^4$He + $^{16}$O, the parabolic 
and the Morse approximations are also inaccurate at energies just above the barrier. It is clear that the results of the Morse approximation are systematically
better than those of the parabola. Nevertheless, they are unsatisfactory, mainly at sub-barrier energies.\\

\begin{figure}%[tb]
\begin{center}
\includegraphics*[width=6.8 cm]{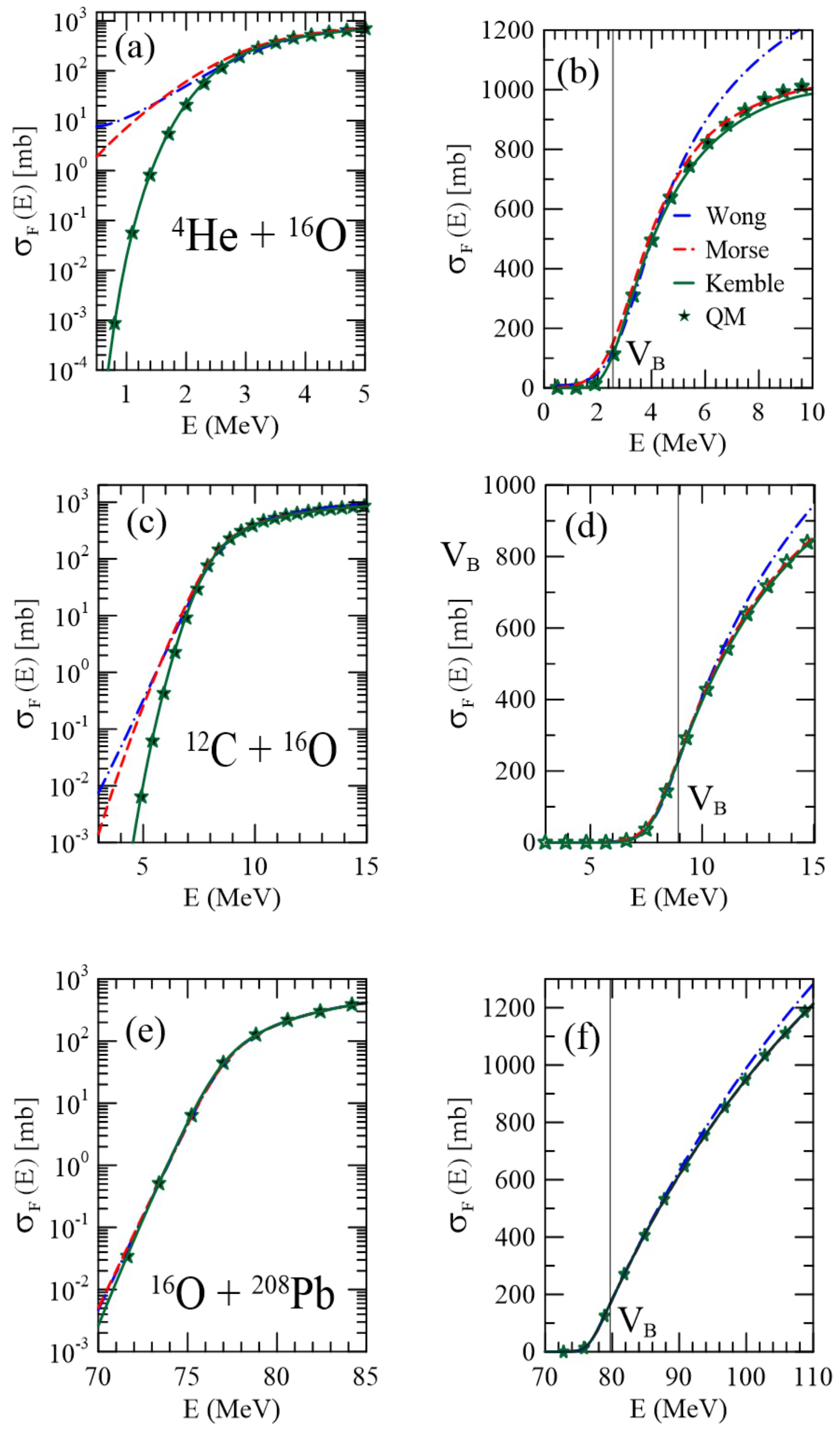}
\end{center}
\caption{(Color on line) Fusion cross sections for the systems of the previous figures. The meaning of the lines are indicated 
within panel (b).
}
\label{sigF}
\end{figure}
Fig.~\ref{sigF} shows the fusion cross sections obtained with different approximations, in comparison with the exact cross sections (stars). The green 
solid lines are the results of Kemble WKB, the blue dot-dashed lines correspond to the Wong formula (Eqs.~(\ref{wong1-a} - \ref{wong1-c}))
and the red dashed lines correspond to the Morse approximation. That is, they are obtained evaluating the partial-wave sum of Eq.~(\ref{lsumsigF}), 
with $P^{\scr F}_l$ approximated by the transmission coefficient through the Morse barrier fitting $V_l(r)$.

\medskip

The main trends of the fusion cross sections of the three systems are similar to the ones observed for the transmission coefficients. First, the Kemble WKB 
reproduces the exact results of the three systems with great accuracy, above and below the Coulomb barrier. Second, the Morse and the Wong cross
sections for the $^{16}$O + $^{208}$Pb at sub-barrier energies reproduce the exact cross section fairly well. On the other hand, the cross sections of the
$^4$He + $^{16}$O and $^{12}$C + $^{16}$O systems obtained with these approximations greatly overestimate the quantum mechanical cross section 
at sub-barrier energies. This is an immediate consequence of the abnormally large transmission coefficients of the parabolic and Morse barriers in this 
energy range. 

\begin{figure}%[tb]
\begin{center}
\includegraphics*[width=7 cm]{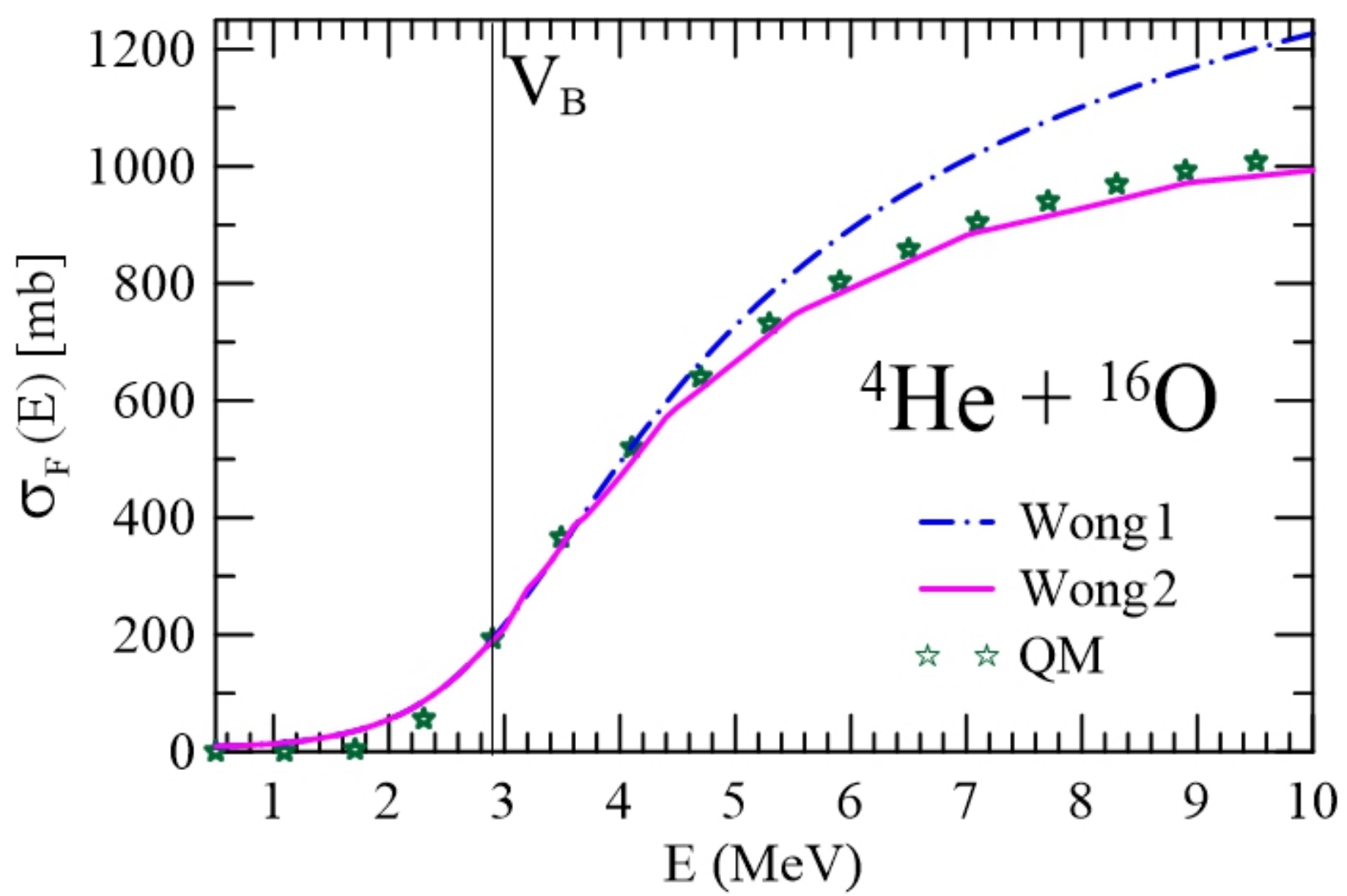}
\end{center}
\caption{(Color on line) Fusion cross sections given by the original Wong (wong 1) formula  and by
the improved Wong formula (wong 2) for the $^4$He + $^{16}$O system, in comparison with the
quantum mechanical results (stars).
}
\label{RH_Wong}
\end{figure}

On the other hand, one can observe an interesting trend of Wong's cross section at energies above the barrier. It is systematically larger than the quantum 
mechanical one. This discrepancy increases as the system's mass decreases. Rowley and Hagino~\cite{RoH15} explained that this is a consequence of using 
a constant barrier radius, independent of the angular momentum.  They pointed out that this problem could be fixed by using the barrier parameters of the 
grazing angular momentum in Wong's formula, as described in Sec.~\ref{Edep_W}. 
This is illustrated in Fig.~\ref{RH_Wong}, where we show the cross section obtained with the original Wong's formula (wong 1) and the ones obtained with 
Wong's formula with energy-dependent parameters (wong 2), as given by Eqs.~(\ref{Nsig} - \ref{Nx})). For comparison, we show also the corresponding
quantum mechanical results (stars). For this illustration, we consider only the $^4$He + $^{16}$O system, where Wong's approximation above the barrier 
is worst. Since the two versions of the Wong formula are identical below the Coulomb barrier, it is not necessary to display the results in a logarithmic scale. 
Inspecting the figure, one concludes that the Wong formula with energy-dependent parameters of Ref.~\cite{RoH15} works very well above the Coulomb
barrier, even in the case of a very light system.

\section{Conclusions}

We have studied the old problem of the one-dimensional tunneling in quantum scattering and fusion. First, we considered approximations of the barrier 
shape, by parabolae (Hill-Wheeler approximation) and by Morse functions. These approximations have the advantage of leading to analytical expressions 
for the tunnelling probabilities. We  investigated also the analytical continuation of the radial variable in Kemble's version of the WKB approximation
for the parabolic and Morse barriers. We have shown that Kemble's WKB on the complex $r$-plane leads to exact transmission coefficients through
these barriers and this conclusion is valid at energies below and above the barrier height.\\

Investigating the S-wave transmission coefficients for the $^4$He + $^{16}$O, $^{12}$C + $^{16}$O and $^{16}$O + $^{208}$Pb systems, we found 
that the Morse barrier, being a more realistic non-symmetric function, leads systematically to better transmission coefficients. However, the improvement 
is not significative, as both approximations are poor at sub-barrier energies, except in the case of $^{16}$O + $^{208}$Pb, or heavier systems. On the other 
hand, the Kemble WKB approximation on the complex $r$-plane using the exact potential, developed in Ref.~\cite{TCH17}, gives a very accurate description of 
the quantum mechanical results, above and below the Coulomb barrier. \\

We performed calculations of fusion cross sections for the above mentioned systems using the Wong, the Morse and Kemble WKB approximations.
The conclusions were similar to the ones reached in the study of transmission coefficients. However, we found that the Wong formula overestimates
the fusion cross section at energies above the barrier, mainly for very light systems. This shortcoming was then eliminated adopting the energy-dependent
Wong formula of Rowley and Hagino~\cite{RoH15}.

\section*{Acknowledgments}
We are grateful Raul Donangelo for critically reading the manuscript. Partial support from the Brazilian funding agencies, CNPq, FAPESP, and FAPERJ 
is acknowledged. M. S. H. acknowledges support from the CAPES/ITA Senior Visiting Professor Fellowship Program.

%%%%%%%%%%%%%%%%%%%%%%%%%%%%%%%%%%%%%%%%%%%%%%%
% \bibliographystyle{apsrev}
% \bibliography{fusbreak-2016_vs03} 
 %%%%%%%%%%%%%%%%%%%%%%%%%%%%%%%%%%%%%%%%%%%%%%%

%\end{document}

%%%%%%%%%%%%%%%%%%%%%%%%%%%%%%%%%%%%%%%%%%%%%%%

%
\end{document}